\documentclass{llncs}  
\usepackage{verbatim}

\usepackage{llncsdoc}
\usepackage{amsmath}
\usepackage{amssymb}

\usepackage{graphicx}

\usepackage{caption}
\usepackage{subcaption}
\usepackage{setspace}

\newcommand{\iseij}[6]{\left\{
    \begin{array}{ccc}
      {#1}&{#2}&{#3}\\
      {#4}&{#5}&{#6}
    \end{array}\right\}}

\title{The Screen representation of spin networks: 2D recurrence,  eigenvalue equation for $6j$ symbols, geometric interpretation and Hamiltonian dynamics.}

\author{Roger W. Anderson\inst{1}\and Vincenzo Aquilanti\inst{2,3}\and Ana Carla P. Bitencourt\inst{4}\and Dimitri Marinelli \inst{5,6} \and Mirco Ragni\inst{4}}

\institute{Department of Chemistry, University of California, Santa Cruz, CA 95064, U.S.A. \email{anderso@ucsc.edu} \and 
Dipartimento di Chimica, Universit\`a di Perugia, Italy \email{vincenzoaquilanti@yahoo.it} \and
Istituto Metodologie Inorganiche e Plasmi CNR, Roma, Italy \and
Departamento de F\'isica, Universidade Estadual de Feira de Santana, Brazil \and
Dipartimento di Fisica, Universit\`a degli Studi  di Pavia, Italy  \and
INFN, sezione di Pavia, 27100 Pavia, Italy}

\begin{document}

\maketitle

\begin{abstract}
This paper treats $6j$ symbols or their orthonormal forms as a function of two variables spanning a square manifold 
which we call the ``screen''. We show that this approach gives important and interesting insight.  This two dimensional perspective provides the most natural extension to exhibit the role of these discrete functions as matrix elements that appear at the very foundation of the modern theory of classical discrete orthogonal polynomials.  Here we present 2D and 1D recursion relations that are useful for the direct computation of the orthonormal $6j$, which we name $U$.  We present a convention for the order of the arguments of the $6j$ that is based on their classical and Regge symmetries, and a detailed investigation of new geometrical aspects of the $6j$ symbols. 
Specifically we compare the geometric recursion analysis of Schulten and Gordon with the methods of this paper.  
The 1D recursion relation, written as a matrix diagonalization problem, permits an interpretation as a discrete Shr\"odinger-like equations and an asymptotic analysis  illustrates semiclassical and classical limits in terms of Hamiltonian evolution.
\end{abstract}

\section{Introduction}\label{sec:s1}

Continuing and extending previous work \cite{bitencourt2012exact,aquilanti2012semiclassical,ragni2010,physreva.58.3718} on $6j$ symbols, (or on the equivalent Racah coefficients), of current use in quantum mechanics and recently also of interest as the elementary building blocks of spin networks \cite{1402-4896-78-5-058103,springerlink:10.1007/s00214-009-0519-y,ac.00}, in this paper we \emph{(i)} - adopt a representation (the ``screen'') accounting for exchange and Regge symmetries; \emph{(ii)} - introduce a recurrence relationship in two variables, allowing not only a computational algorithm for the generation of the $6j$ symbols to be plotted on the screen, but also representing a partial difference equation allowing us to interpret the dynamics of the two dimensional system. \emph{(iii)} - introduce a recurrence relationship as an equation in one variable, extending the known ones which are also computationally interesting; \emph{(iv)} - give a formulation of the difference equation as a matrix diagonalization problem, allowing its interpretation as a discrete Schr\"odinger equation; \emph{(v)} - discuss geometrical and dynamical aspects  from an asymptotic analysis. We do not provide here detailed proofs of these results, but give sufficient hints for the reader to work out the derivations. For some of the topics we refer to a recent problem recently tackled \cite{aquilanti2013volume}; numerical and geometrical illustrations are presented on a companion paper \cite{lncs2.2013}. A concluding section introduces aspects of relevance for the general spin networks by sketching some features of the $9j$ symbols.

\section{The screen: Classical and Regge symmetries, Canonical Form}\label{s4}

The Wigner $6j$ symbols $\iseij{j_1}{j_2}{j_{12}}{j_3}{j}{j_{23}}$ are defined as a matrix element beetween alternative angular momentum coupling schemes \cite{varsh} by the relation
\begin{equation}
\left\langle j_1j_2\left( j_{12}\right) j_3jm\mid j_1 j_2j_3\left( j_{23}\right) j'm'\right\rangle = \\
\left( -1\right) ^{j_1+j_2+j_3+j} \delta_{jj'}\delta_{mm'}U\left( j_1j_2jj_3;j_{12}j_{23}\right)\nonumber ,
\end{equation}
where the orthonormal transformation $U$ is
\begin{equation}\label{eq:uj12}
U\left( j_1j_2jj_3;j_{12}j_{23}\right)=\sqrt{\left( 2j_{12}+1\right) \left( 2j_{23}+1\right)}\\
\iseij{j_1}{j_2}{j_{12}}{j_3}{j}{j_{23}} 
\end{equation}
For given values of $j_1$, $j_2$, $j_3$, and $j$ the $U$ will be defined over a range for
both $j_{12}$ and $j_{23}$.  These ranges are given by \\

\begin{alignat}{6}
j_{12~min} & = & \max\left( \mid j_1-j_2 \mid, \mid j-j_3\mid\right), \quad%
& j_{12~max} & = &\min\left( j_1+j_2, j+j_3\right),\nonumber\\%
j_{23~min}&=&\max\left( \mid j_1-j \mid, \mid j_2-j_3\mid\right),\quad  %
&j_{23~max}&=&\min\left( j_1+j, j_2+j_3\right), \nonumber \\%
\mbox{and }& &j_{12~min}\leq j_{12} \leq j_{12~max},   ~~~ & j_{23~min}\leq & j_{23}& \leq j_{23~max} .\label{eq:Rangej12j23}
\end{alignat}  
The screen corresponds to the $6j$ or, as we specify below, the $U$ values for all possible values of $j_{12}$ and $j_{23}$ .

The range for $j_{12}$ and $j_{23}$ is determined by the values of the independent variables:  
$j_1$, $j_2$, $j_3$, and $j$.  In the remainder of this paper we make this clear by introducing new variables $a,\, b,\ c,\ d,\ x$ and $y$  to replace the $j$ values.  We specify the new variables by establishing a correspondence:
\begin{equation}\label{eq:abx}
\iseij{a}{b}{x}{c}{d}{y} \Leftrightarrow \iseij{j_1}{j_2}{j_{12}}{j_3}{j}{j_{23}}
\end{equation}
Assuming that $x$ and $y$ remain respectively in the upper and lower right side of the $6j$ symbols, there are four  classical and one Regge relevant symmetries:
\begin{equation}\label{eq:symm}
\iseij{a}{b}{x}{c}{d}{y} = \iseij{b}{a}{x}{d}{c}{y} =\iseij{d}{c}{x}{b}{a}{y} =\iseij{c}{d}{x}{a}{b}{y}\\
=\iseij{s-a}{s-b}{x}{s-c}{s-d}{y}, 
\end{equation}
where $s=\left( a+b+c+d\right)/2 $ .
It can be shown \cite{aquilanti2012semiclassical,bitencourt2012exact} that $x_{max}-x_{min} = y_{max}-y_{min} = 2\min \left( a,b,c,d,s-d,s-c,s-b,s-a \right) = 2\kappa$.  The square screen will contain $\left( 2\kappa+1\right) ^2 $ values.
The canonical ordering for $6j$ screens can now be specified by considering the two sets of values: $a$, $b$, $c$, $d$ and its Regge transform $a'=s-a$, $b'=s-b$, $c'=s-c$, and $d'=s-d$. Take the set with the smallest entry and use the classical $6j$ symmetries to place this smallest value in the upper left corner of the $6j$ symbol.  The placement of the other $6j$ arguments are determined by the symmetry relations. The resulting symbol has the property that $x_{\min}=b-a \leq x \leq b+a = x_{\max}$ and $y_{\min}=d-a \leq y \leq d+a = y_{\max}$.  Furthermore we require that $a \leq b \leq d$ for the Canonical form.  This may require using Eq. \ref{eq:Can6j} to "orient" the screen in this way.  
\begin{gather}\label{eq:Can6j}
	\iseij{a}{b}{x}{c}{d}{y}=	\iseij{a}{d}{y}{c}{b}{x}
\end{gather}
It can be shown that any symbol to be studied as a function of two entries can be reduced to the canonical form of Eq.  \ref{eq:Can6j} where 
$a\leq b \leq d \leq b+c-a$ and $c_{\min}=d-a+b \leq c \leq d+a-b=c_{\max}$.

Regge transformation for the parameters of the screen is a linear 
$O(4)$ transformation:
\begin{gather}\label{eq:defR}
\frac{1}{2} 
\begin{pmatrix}
-1 & 1 & 1 & 1\\
1 & -1 & 1 & 1\\
1 & 1 & -1 & 1\\
1 & 1 & 1 & -1
\end{pmatrix}
\begin{pmatrix}
a\\
b\\
c\\
d
\end{pmatrix}
=
\begin{pmatrix}
s-a\\
s-b\\
s-c\\
s-d
\end{pmatrix}.
\end{gather}
It can be checked that several functions appearing below (caustics, ridges, etc.) are invariant under such symmetry and also when represented on the screen (See \cite{lncs2.2013}).

\section{Tetrahedra and $6j$ symbols}\label{s3}

\begin{figure}
      \centering
      \begin{minipage}[b]{0.30\textwidth}
         \centering
         \includegraphics[width=\textwidth]{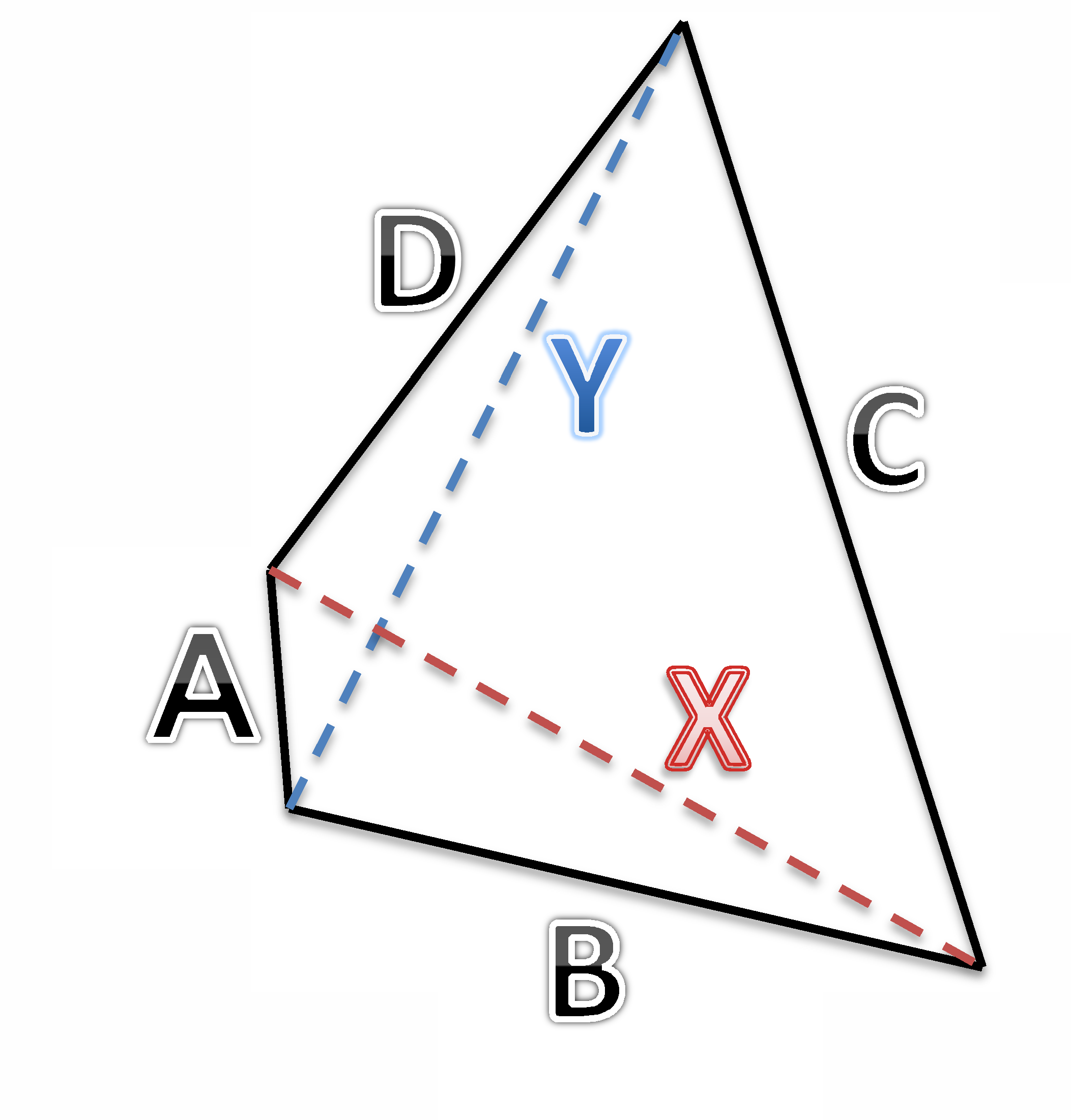}
         \subcaption{Ponzano-Regge tetrahedron built with the six angular momenta in the $6j$ symbol.}
         \label{fig:tetra}
      \end{minipage}%
     ~~ \begin{minipage}[b]{0.60\textwidth}
         \centering
         \includegraphics[width=\textwidth]{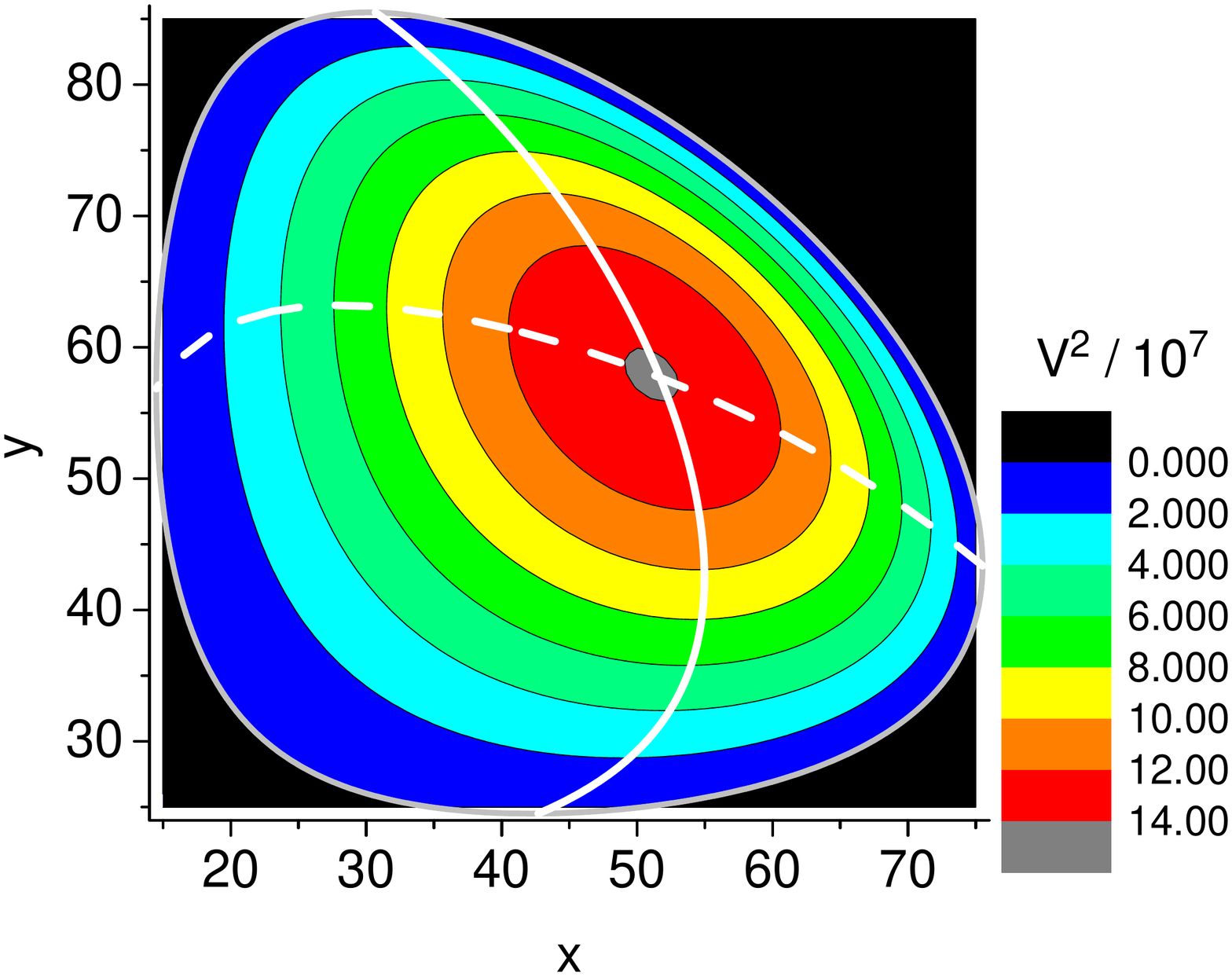}
         \subcaption{$V^2$ (contours for Eq. \ref{eq:volume}), caustics Eq. \ref{eq:yvzero} (gray boundary), ridges (solid white Eq. \ref{eq:yvmax}, dashed Eq. \ref{eq:xvmax}) for $a = 30$, $b = 45$, $c = 60$, and $d = 55$.}
         \label{fig:ridges}
      \end{minipage}  
    \caption{}
\end{figure}

In the following when we consider $6j$ properties as correlated to
those of the tetrahedron of Figure \ref{fig:tetra} \cite{ragni2010}, we use the
substitutions $A = a + 1/2$, $B = b + 1/2$, $C = c + 1/2$, $D = d + 1/2$, $X = x + 1/2$, $Y = y + 1/2$ which greatly improves all
asymptotic formulas down to surprisingly low values of the
entries.  We show the argument ranges where the correspondence with the tetrahedron breaks down in section \ref{sec:geo}.

The area of each triangular face is given by the
Heron formula:
\begin{equation}\label{eq:area}
F(A,B,C) = \frac{1}{4}\sqrt{(A+B+C)(-A+B+C)(A-B+C)(A+B-C)}
\end{equation}
 where $A$, $B$, $C$ are the sides of the face.  Upper case letters are used here to stress that geometric lengths are used in the equation.  The square of the area can be also expressed as a Cayley-Menger determinant.   Similarly, the square of the volume of an irregular tetrahedron, can also be written as a Cayley-Menger determinant (Eq. \ref{eq:volume} or as a 
Gramian determinant \cite{fl03}.  The latter determinant 
embodies a clearer relationship with a vectorial picture but with partial spoiling of the symmetry. 

\begin{equation}\label{eq:volume}
V^2 = \frac{1}{288}
\left| \begin{array}{ccccc}
    0        & C^2  & D^2      & Y^2 & 1\\
    C^2      & 0    & X^2      & B^2 & 1\\
    D^2      & X^2  & 0        & A^2 & 1\\
    Y^2      & B^2  & A^2      & 0   & 1\\
    1        & 1    & 1        & 1   & 0
  \end{array}\right|.
\end{equation}

An explicit formula, due to Piero della Francesca,  will be used in the companion paper \cite{lncs2.2013}.
Additionally mirror symmetry \cite{bitencourt2012exact}, can be used to extend screens to cover a larger range of arguments.  The appearance of squares of
tetrahedron edges entails that the invariance with respect to the
exchange $X \leftrightarrow -X$ implies formally $x
\leftrightarrow -x-1$ with respect to entries in the $6j$
symbol. Although this is physically irrelevant when the $j$'s are
pseudo-vectors, such as physical spins or orbital angular momenta,
it can be of interest for other (e.g. discrete algorithms)
applications. Regarding the screen, it can be seen that actually
by continuation of $X$ and $Y$ to negative values, one can have
replicas that can be glued by cutting out regions shaded in Fig. …
in \cite{jpa2012}, allowing mapping onto the $S^2$ manifold.

Figure \ref{fig:ridges} 
illustrates $V^2$ for values of $a, b, c$, and $d$ used later in this paper. \\

The following equations were first introduced in Refs. \cite{bitencourt2012exact} and \cite{ragni2010}, but they are rewritten here with changed notation. When the values of $A$, $B$, $C$, $D$ and $X$ are fixed, the maximum value
for the volume as a function of $Y$ is given by the ``ridge'' curve
 \begin{equation}\label{eq:yvmax}
  Y^{Vmax} = \left(\frac{(A^2-B^2)(C^2-D^2)+
  (A^2+B^2+C^2+D^2)X^2-X^4}{2X^2}\right)^{1/2}~,
 \end{equation}
 the corresponding volume is
 \begin{equation}\label{eq:vyvmax}
  V^{max}(A, B, C, D, X)=\frac{\sqrt{\Lambda_{A,B,X}\Lambda_{C,D,X}}}{24X}~,
 \end{equation}
 where
 \begin{equation}\label{eq:lambda}
 \Lambda_{\alpha,\beta,\gamma} = \left(\alpha^2 -\beta^2\right)^2-2\gamma^2\left (\alpha^2+\beta^2\right ) + \gamma^4 .
  \end{equation}
 Therefore the two values of $Y$ for which the volume is zero are:
\begin{equation}\label{eq:yvzero}
 Y^z = \left(\left(Y^{Vmax}\right)^2\pm\frac{\sqrt{\Lambda_{A,B,X}\Lambda_{C,D,X}}}{2X^2}\right)^{1/2}.
\end{equation}
The values for $Y^z$ mark the boundaries between classical and nonclassical
regions, and therefore called ``caustics''.

 Also when the values of
 $A$, $B$, $C$, $D$ and $Y$ are fixed, the maximum value
 for the volume as a function of $X$ is given by the other ``ridge'' curve:
 \begin{equation}\label{eq:xvmax}
  X^{Vmax} = \left(\frac{(A^2-D^2)(C^2-B^2)+
  (A^2+B^2+C^2+D^2)Y^2-Y^4}{2Y^2}\right)^{1/2}~.
 \end{equation}

\section{Recursion formulas and exact calculations}\label{sec:s5}

The $U$ values that are represented on  the screen must be calculated by efficient and accurate algorithms, and we employed  several  methods that we have previously discussed and tested. Explicit formulas are available either as sums over a single
variable and series, and we have used such calculations with multiple precision arithmetic in previous work \cite{307},\cite{ragni2010}, \cite{AnAqMa}, \cite{aaf.08} .  These high accuracy calculations are entirely reliable for all $U$ that we have considered in the past, and the results provide a stringent test for other methods. However recourse to recursion formulas appears most convenient for fast accurate calculations and -as we will
emphasize- also for semiclassical analysis, in order to understand
high $j$ limit and in reverse to interpret them as discrete
wavefunctions obeying Schr\"odinger type of difference (rather than
differential) equations. 

The goal is to determine the elements of the ortho-normal transformation matrix:
\begin{equation}\label{defu}
U\left( x,y \right) =\sqrt{\left(2x+1 \right)\left(2y+1 \right)}\iseij{a}{b}{x}{c}{d}{y}.
\end{equation}
Two approaches can be used to evaluate ${U}\left( x,y\right)$:  evaluate the $6j$ from recursion formulas and then apply the normalization or to use direct calculation from explicit formulas.

\subsection{2D ($x$,$y$) recursion for $U$}
In this work, we first derive and 
computationally implement a two variable recurrence that permits
construction of the whole orthonormal matrix The derivation
follows our paper in \cite{AnAqMa} and is also
of interest for other $3nj$ symbols. 

By setting $h = 0$ in the formula \ref{eq:recur9j} in section \ref{sec:9j} , we obtain a five term recurrence relation for $U(x,y)$:

\begin{eqnarray}\label{eq:xyrecur}
\left( { - 1} \right)^{2x }\sqrt {\frac{{2x - 1}}{{2y + 1}}} 
  \iseij{b}{x-1}{a}{1}{\:a\:}{x}
  \iseij{d}{x-1}{c}{1}{\:c\:}{x}
  U\left( {x - 1,y} \right) \nonumber \\
 + {\left( { - 1} \right)^{2x}}\sqrt {\frac{{2x + 1}}{{2y + 1}}} 
  \iseij{b}{x}{a}{1}{a}{x}
  \iseij{d}{x}{c}{1}{c}{x}
  U\left( {x,y} \right) \nonumber \\
 + {\left( { - 1} \right)^{2x }}\sqrt {\frac{{2x + 3}}{{2y + 1}}} 
  \iseij{b}{x+1}{a}{1}{a}{x}
  \iseij{d}{x+1}{c}{1}{c}{x}
  U\left( {x + 1,y} \right) \nonumber \\
= {\left( { - 1} \right)^{2y }}\sqrt {\frac{{2y - 1}}{{2x + 1}}} 
  \iseij{b}{y-1}{c}{1}{c}{y}
  \iseij{d}{y-1}{a}{1}{a}{y}
  U\left( {x,y-1} \right) \nonumber \\
+ {\left( { - 1} \right)^{2y}}\sqrt {\frac{{2y + 1}}{{2x + 1}}} 
  \iseij{b}{y}{c}{1}{c}{y}
  \iseij{d}{y}{a}{1}{a}{y}
  U\left( {x,y} \right) \nonumber \\
+ {\left( { - 1} \right)^{2y }}\sqrt {\frac{{2y + 3}}{{2x + 1}}}
  \iseij{b}{y+1}{c}{1}{c}{y}
  \iseij{d}{y+1}{a}{1}{a}{y}
  U\left( {x,y+1} \right)
\end{eqnarray}

This recurrence relation Eq. \ref{eq:xyrecur} will yield the entire set of $U(x,y)$ that constitute the screen.  Replacing the $6j$ symbols of unit argument with the algebraic expressions in Varshalovich \cite{varsh}, we obtain an effective method to calculate the screen.   \\

\section{1D ($x$) symmetric recursion for $U$} \label{sec:s6}

Starting with the recurrence relation in Neville \cite{NEV} and Schulten and Gordon \cite{schgorb} for the $6j$ and carefully converting it into a recurrence relation for $U$, we can write a three term symmetric recursion relationship, which is here conveniently represented as an eigenvalue equation:

\begin{equation}\label{eq:xrecur}
{p_ +} \left(x\right) U\left( {x + 1,y} \right) + w(x)U\left( {x,y} \right) + {p_ - }\left(x\right)U\left( {x - 1,y} \right) = \lambda \left( y \right)U\left( {x,y} \right)~,
\end{equation}
where

\begin{equation}\label{eq:gpexact}
\begin{split} 
{p_ + }\left( x \right) = {\left\{ {\left( {a + b + x + 2} \right)\left( {a + b - x} \right)\left( {a - b + x + 1} \right)\left( { - a + b + x + 1} \right)} \right\}^{\frac{1}{2}}}  \\
 \times {\left\{ {\left( {d + c + x + 2} \right)\left( {d + c - x} \right)\left( {d - c + x + 1} \right)\left( { - d + c + x + 1} \right)} \right\}^{\frac{1}{2}}}  \\
 \times {\left( {x + 1} \right)^{ - 1}}{\left[ {\left( {2x + 1} \right)\left( {2x + 3} \right)} \right]^{ - \frac{1}{2}}} \\
\end{split}
\end{equation}
\begin{equation}\label{eq:gmexact}
{p_ - }\left( x \right) = {{ p}_ + }\left( {x - 1} \right)\\
\end{equation}
\begin{equation}\label{eq:g0exact}
\begin{split}
{w}\left( x \right) =   \left[ {b\left( {b + 1} \right) - a\left( {a + 1} \right) + x\left( {x + 1} \right)} \right] \\
 \times \left[ {d\left( {d + 1} \right) - c\left( {c + 1} \right) - x\left( {x + 1} \right)} \right]/\left[ {x\left( {x + 1} \right)} \right]\\
\end{split}
\end{equation}
\begin{equation}\label{eq:eigen}
\lambda \left( y \right) = 2\left[ {y\left( {y + 1} \right) - b\left( {b + 1} \right) - c\left( {c + 1} \right)} \right]~.
\end{equation}

For convenience we can also define:
\begin{equation}
w_{\lambda} = w\left(x\right) - \lambda \left(y\right)
\end{equation}

A row of the screen may be efficiently and accurately calculated from these equations.  Diagonalization of the symmetric tridiagonal matrix given by the $p_+\left(x\right),\ w(x),\ p_-\left(x\right)$ provides an accurate check: the eigenvalues of the tridiagonal matrix precisely match those expected from Eq. \ref{eq:eigen} and eigenvectors generate $U(x,y)$ . Stable results are obtained with double precision arithmetic.\\

\subsection{Potential functions and Hamiltonian dynamics}

\begin{figure}
      \centering
      \begin{minipage}[b]{0.48\textwidth}
         \centering
         \includegraphics[width=\textwidth]{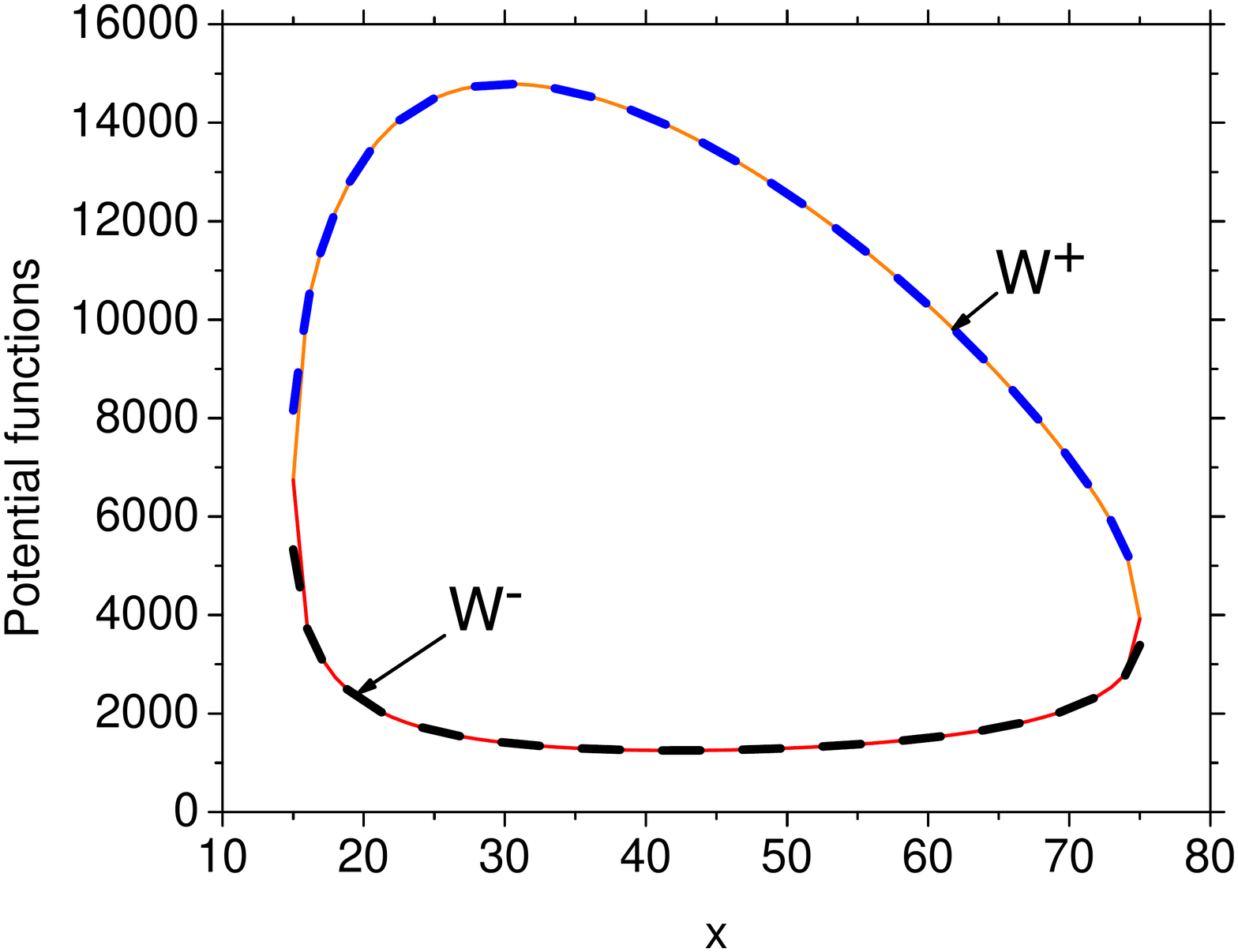}
         \subcaption{Angular momenta corresponding to Figure \ref{fig:ridges}}
         \label{fig:braunpot_1a}
      \end{minipage} %
    ~~ \begin{minipage}[b]{0.48\textwidth}
         \centering
         \includegraphics[width=\textwidth]{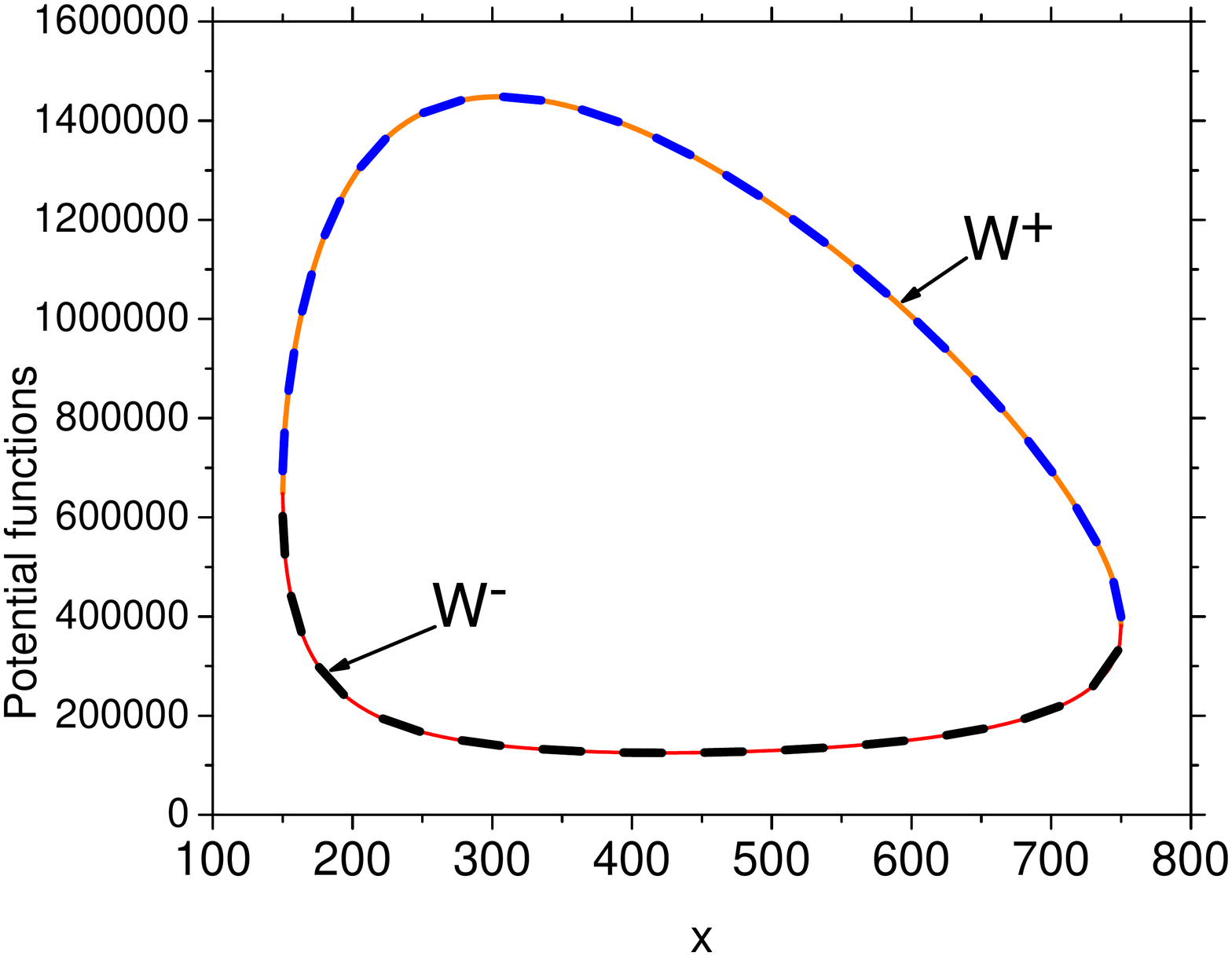}
         \subcaption{Angular momenta: $a = 300$, $b = 450$, $c = 600$, and $d = 550$.}
         \label{fig:braun_1a_10x}
      \end{minipage}  
    \caption{Potential functions corresponding to Eq. \ref{eq:braun_arit} (dashed blue and black lines) and Eq.\ref{eq:braun_geo} (thin solid orange and red lines)}
\end{figure}

For the eigenvalue equation (Eq. \ref{eq:xrecur}), interpreted as discrete Schr\"odinger-like equation,  two potentials $\textbf{W}^+\left(x\right)$ and 
$\textbf{W}^-\left(x\right)$ can be defined: 
\begin{equation}\label{eq:braun}
 \textbf{W}^\pm\left(x\right)= w(x)  \pm 2\mid  {p}\left(x\right) \mid ,
\end{equation}
where \cite{braun1993discrete}
\begin{equation}\label{eq:braun_arit} 
 {p}\left(x\right) = \frac{1}{2} \left({p_+\left(x\right)+p_-\left(x\right)}\right) \end{equation}
or \cite{braun78a}
\begin{equation}\label{eq:braun_geo}
 {p}\left(x\right) =  \sqrt{\left({p_+\left(x\right)p_-\left(x\right)}\right)}.  
\end{equation}
The two definitions agree well except for $x$ near the limits $x_{min}$ or $x_{max}$.  With the second choice for $\bar{p}\left(x\right)$ the values for  $\textbf{W}^\pm$ are the same at the limits, but there are differences with the first choice.    See the figures \ref{fig:braunpot_1a} and \ref{fig:braun_1a_10x}.  Compare with Ref. \cite{aquilanti2013volume} where Hamiltonian dynamics is developed for a similar system.  Braun's potential functions are closely related to the caustics illustrated in \cite{bitencourt2012exact} and \cite{lncs2.2013}.

\subsection{Geometric interpretation}\label{sec:geo}
The geometrical interpretations of the $6j$ symbols provide  fundamental understanding and important semiclassical limits.  This
approach originates from Ponzano and Regge \cite{ponzregge} and
elaborated by others, notably Schulten and Gordon \cite{schgorb}.

The three-term recursion relationship (Eq. \ref{eq:xrecur}), for
$U$ admits an illustration in terms of a geometric
interpretation: with some approximations to be detailed below one
has finite difference equations (see Ref.\cite{NEV}, Eq.(67) for
relationships between recursions and finite difference), Consider
the Schulten-Gordon relationships Eq.(66) and Eq.(67)(Ref. \cite{schgorb}).  
Here we show new geometric representations of the recursion relationships.

By setting $a=A-\tfrac{1}{2}$, $b=B-\tfrac{1}{2}$, $c=C-\tfrac{1}{2}$, $d=D-\tfrac{1}{2}$, $x=X-\tfrac{1}{2}$, and $y=Y-\tfrac{1}{2}$ one can write Eq. \ref{eq:xrecur} in terms of triangle areas, a length $X'$, and the cosine of a dihedral angle $\theta_3$.  The accuracy of this approximation is excellent, and depends slightly on the choice for $X'$.

\begin{eqnarray}\label{eq:bestapprox}
 \frac{F(X-\frac{1}{2},A,B)F(X-\frac{1}{2},C,D)}{\left( X-\frac{1}{2}\right) ^2}U\left(x-1,y\right)\nonumber\\
  +\frac{F(X+\frac{1}{2},A,B)F(X+\frac{1}{2},C,D)}{\left( X+\frac{1}{2}\right) ^2}U\left(x+1,y\right)\nonumber\\
  -2\cos\theta_{3}\frac{F(X',A,B)F(X',C,D)}{X'^2}U\left(x,y\right)\approx 0
\end{eqnarray}
and Eq.(69)\cite{schgorb} {\footnotesize \begin{equation}\label{eq:PRSG2a}
  \cos\theta_{3}=\frac{2X'^2Y^2-X'^2\left( -X'^2+D^2+C^2\right)-B^2\left(X'^2+D^2-C^2\right)-A^2\left(X'^2-D^2+C^2\right)}{16F\left(X',B,A\right)F\left(X',D,C\right)},
\end{equation}}
where $F(a,b,c)$ is ``area" of $abc$ triangle (Eq. \ref{eq:area}).  (This recursion relation Eq. \ref{eq:bestapprox} must be multiplied through by $8$ to compare precisely with Eq. \ref{eq:xrecur}.

\begin{figure}
      \centering
      \begin{minipage}[b]{0.48\textwidth}
         \centering
         \includegraphics[width=\textwidth]{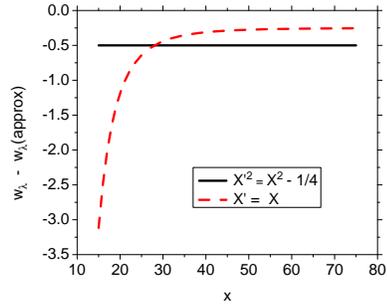}
         \subcaption{Error in $w_\lambda$}
         \label{fig:errg01a}
      \end{minipage} %
     ~~ \begin{minipage}[b]{0.48\textwidth}
         \centering
         \includegraphics[width=\textwidth]{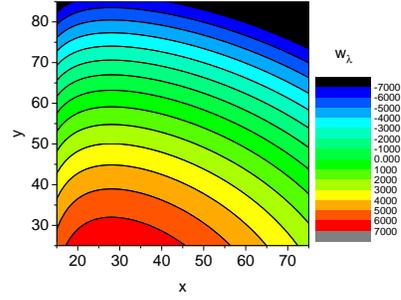}
         \subcaption{$w_\lambda$ values}
         \label{fig:g0fig01a}
      \end{minipage}  
    \caption{Parameters $a,b,c,d$ of Figure \ref{fig:ridges}}
\end{figure}

Here we consider two choices for $X'$ in Eq. \ref{eq:g0exact}:
\begin{equation}\label{eq:xprime}
X'^2 = \left(X-\frac{1}{2}\right)\left(X+\frac{1}{2}\right)= X^2-\frac{1}{4},
\end{equation}
\begin{equation}\label{eq:xisx}
X' = X
\end{equation}
The first choice (Eq. \ref{eq:xprime}) provides an almost exact approximation to Eqns. \ref{eq:gpexact}, \ref{eq:gmexact},\ref{eq:g0exact}, and \ref{eq:eigen}
Coefficients in Eq. \ref{eq:xrecur}.  The second  Eq. \ref{eq:xisx} uses only integer or half integer arguments, and for most $X$ works as well as the first. 
The figures \ref{fig:errg01a} and \ref{fig:g0fig01a} show the errors and their significance.  In these figures $w_\lambda\left(approx\right)$ is specified as:
\begin{equation}
w_\lambda\left(approx\right) = -2\cos\theta_{3}\frac{F(X',A,B)F(X',C,D)}{X'^2}
\end{equation}

  For either choice of $X'$, the recursion coefficients are connected to the geometry of tetrahedra \cite{ponzregge}:
\begin{equation}\label{eq:PRSG3}
 \frac{3}{2}VX'=F(X',A,B)F(X',C,D)\sin\theta_{3}~,
\end{equation}
where $V$ is the tetrahedral volume. 

Equations \ref{eq:bestapprox} can be recast by the geometric mean approximation:  
\begin{equation}\label{eq:sgapprox}
 \frac{F(X\pm\frac{1}{2},A,B)}{\left( X\pm\frac{1}{2}\right)}\simeq
 \frac{\sqrt{F(X\pm1,A,B)F(X,A,B)}}{\sqrt{X\left( X\pm1\right)}}, 
\end{equation}
where $A$ and $B$ can be also replaced by $C$ and $D$.

With Eq. \ref{eq:sgapprox}  Eq. \ref{eq:bestapprox} becomes:
\begin{eqnarray}\label{eq:oneapprox}
 \frac{\sqrt{F(X-1,A,B)F(X,A,B)F(X-1,C,D)F(X,C,D)}}{X\left( X-1\right)} U\left(x-1,y\right)\nonumber\\
  + \frac{\sqrt{F(X+1,A,B)F(X,A,B)F(X+1,C,D)F(X,C,D)}}{X\left( X+1\right)} U\left(x+1,y\right)\nonumber\\
  -2\cos\theta_{3}\frac{F(X,A,B)F(X,C,D)}{X^2}U\left(x,y\right)\approx 0,
\end{eqnarray}

\begin{figure}
      \centering
      \begin{minipage}[b]{0.48\textwidth}
         \centering
         \includegraphics[width=\textwidth]{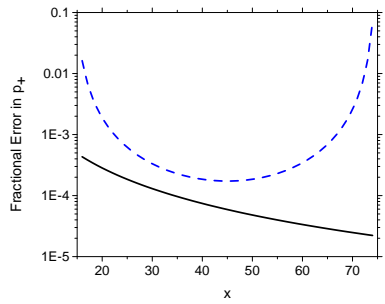}
         \subcaption{Angular momenta of figure \ref{fig:ridges}}
         \label{fig:errgp1a}
      \end{minipage} %
     ~~ \begin{minipage}[b]{0.48\textwidth}
         \centering
         \includegraphics[width=\textwidth]{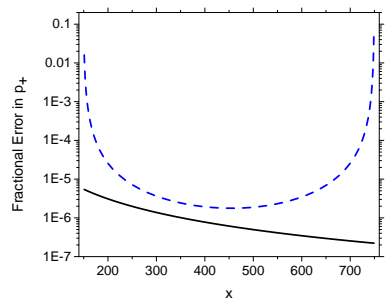}

         \subcaption{Angular momenta of figure \ref{fig:braun_1a_10x} }
         \label{fig:errgp1b}
      \end{minipage}  
    \caption{Fractional errors in $p_+$ using Eq. \ref{eq:bestapprox}, solid black line and Eq. \ref{eq:oneapprox}, dashed blue line. }

\end{figure}

This equation is useful, but definitely less accurate than Eq. \ref{eq:bestapprox} (See Figures \ref{fig:errgp1a} and \ref{fig:errgp1b}).

With cancellation of terms in $X$, this Eq. \ref{eq:oneapprox} becomes:
\begin{eqnarray}\label{eq:sgapprox1}
 \frac{\sqrt{F(X-1,A,B)F(X-1,C,D)
 }}{\left( X-1\right)} U\left(x-1,y\right)\nonumber\\
  + \frac{\sqrt{F(X+1,A,B)F(X+1,C,D)}}{\left( X+1\right)} U\left(x+1,y\right)\nonumber\\
  -2\cos\theta_{3}\frac{\sqrt{F(X,A,B)F(X,C,D)}}{X}U\left(x,y\right)\approx 0.
\end{eqnarray}

This is equivalent to the recursion relation of Schulten and Gordon \cite{schgorb}, that they use to establish their semiclassical approximations for $6j$ symbols.  Their equation is accurate enough for $x_{min} \ll x \ll x_{max}$, but not so accurate near the limits. 

In terms of the finite difference operator, Eq. \ref{eq:oneapprox} becomes after using Eq. \ref{eq:PRSG3}:
$\Delta^2(x)f(x)=f(x+1)-2f(x)+f(x-1)$:
\begin{equation}\label{eq:PRSG4}
 [\Delta^2(X)+2-2\cos\theta_{3}]f(X)\simeq 0~,
\end{equation}
where
\begin{equation}\label{eq:PRSG5}
  f(X)=\frac{\sqrt{F(X,A,B)F(X,C,D)}}{X}U\left(x,y\right)=\sqrt{\frac{V}{X\sin\theta_{3}}}U\left(x,y\right)~.
\end{equation}
We have, explicitly
\begin{equation}\label{eq:PRSG6}
 \cos\theta_{3}=\pm\sqrt{1-\left(\frac{3VX}{2F(X,A,B)F(X,C,D)}\right)^2}~.
\end{equation}

Our Eq. \ref{eq:PRSG5} is only slightly different from that of Schulten and Gordon, because we have an extra $X$ in the denominator of the definition of $f\left(X\right)$.  This occurs because we use the recursion for $U$ instead of that for $6j$.  

\subsection{Semiclassical approximation}\label{sec:semi}

The following developments parallel those in \cite{ragni2010}. From the above formulas, and from that of the volume, we have that
\begin{itemize}
    \item $V=0$ implies $\cos\theta_{3}=\pm 1$ and establishes the classical domain between $X_{min}$ and $X_{max}$
    \item $F(X,A,B)=0$ or $F(X,C,D)=0$ establish the definition limits $x_{min}$ and $x_{max}$. 
\end{itemize}

\begin{figure}
\centering
  \includegraphics[width=0.6\textwidth]{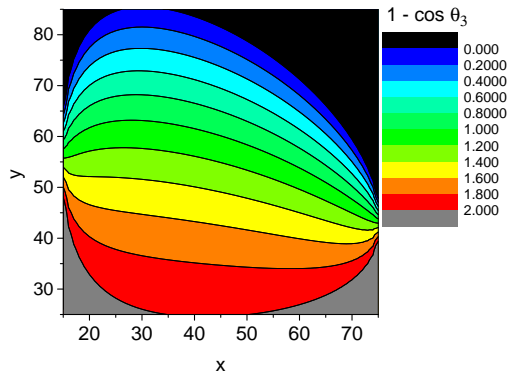}\\
  \caption{x,y plot of cos $\theta_3$ for angular momenta of Fig 1(b)}\label{fig:cost31a}
\end{figure}

For a Schr\"odinger type equation
\begin{equation}\label{eq:PRSG7}
 \frac{d^2\psi}{dx^2}+p^2\psi=0~~,~~~~~~~~~~\hbar^2/2m=1~,
\end{equation}
its discrete analog in a grid having one as a step,
\begin{equation}\label{eq:PRSG8}
 \psi_{n+1} + (p^2-2)\psi_n + \psi_{n-1}=0,
\end{equation}
and we then have after comparing Eq. \ref{eq:PRSG8} with Eq. \ref{eq:PRSG4}
\begin{equation}\label{eq:PRSG9}
 f(X+1)-2\cos\theta_{3}f(X)+f(X-1)=0.
\end{equation}
The identification
\begin{equation}\label{eq:PRSG10}
 p=\pm (2-2\cos\theta_{3})^{1/2}
\end{equation}
is then evident.
Here we present a x,y plot Fig. \ref{fig:cost31a} of $1-\cos\theta_3$ that clearly shows this definition of the classical region.

Evidentially, on the closed loop, we can enforce Bohr-Sommerfeld phase space quantization:

\begin{equation}\label{eq:PRSG11}
 \oint p\,{\rm d}x=\left(n+1/2\right)\,\pi~.
\end{equation} 

The eigenvalues $n$ obtained in this way may be easily related to the allowed $y$.  
These formulas are illustrated in Fig. 6 and 7 of Ref \cite{ragni2010}.

The Ponzano-Regge formula for the $6j$ in the classical region is

\begin{equation} \label{eq:PRs}
\iseij{a}{b}{x}{c}{d}{y} \approx \frac{1}{\sqrt{12 \pi \vert V \vert}} \cos\left(\Phi\right),
\end{equation}
where the Ponzano-Regge phase is:  
$\Phi = A\theta_1+B\theta_2+X\theta_3+C\eta_1+D\eta_2+Y\eta_3+\frac{\pi}{4}$.  The angles are determined by rearranged equations: Eq. \ref{eq:PRSG3}.  The various dihedral angles are found from the equations in \cite{ponzregge}.

To the extent that the Ponzano-Regge approximation is valid we see that the $6j$ symbols have a magnitude envelop given by $V$ and a phase that is a function of $X$ and $Y$ determined by $\Phi$.  Eq. \ref{eq:PRs} works quite well for $X$ and $Y$ well within the classical region.  However its use near the caustics is limited because of two factors:
\begin{enumerate}
\item
The approximate recursion relation given by Eq. \ref{eq:oneapprox} differs most from the exact recursion Eqs. \ref{eq:gpexact},\ref{eq:gmexact},\ref{eq:g0exact},\ref{eq:eigen} near the caustics.
\item
The semiclassical approximation for the $6j$ also breaks down near the caustics.
\end{enumerate}
For piece-wise extensions , see \cite{ponzregge} and for uniformly valid formulas see \cite{schgorb}.

\section{$9j$ and higher spin networks}\label{sec:9j}
In this work, we first have derived and 
computationally implemented a two variable recurrence that permits
construction of the whole orthonormal matrix The derivation
follows our paper in \cite{AnAqMa} and is also
of interest for other $3nj$ symbols. 

We find in \cite{AnAqMa}; see also \cite{varsh}, the following 2D recurrence relationship for $9j$ symbols:

\begin{flalign}\label{eq:recur9j}
\begin{array}{l}
\frac{{{A_{c + 1}}\left( {ab,fj} \right)}}{{\left( {c + 1} \right)\left( {2c + 1} \right)}}\left\{ {\begin{array}{*{20}{c}}
a&b&{c + 1}\\
d&e&f\\
g&h&j
\end{array}} \right\} + \frac{{{A_c}\left( {ab,fj} \right)}}{{c\left( {2c + 1} \right)}}\left\{ {\begin{array}{*{20}{c}}
a&b&{c - 1}\\
d&e&f\\
g&h&j
\end{array}} \right\} - \frac{{{A_{d + 1}}\left( {ef,ag} \right)}}{{\left( {d + 1} \right)\left( {2d + 1} \right)}}\left\{ {\begin{array}{*{20}{c}}
a&b&c\\
{d + 1}&e&f\\
g&h&j
\end{array}} \right\}\\
 - \frac{{{A_d}\left( {ef,ag} \right)}}{{d\left( {2d + 1} \right)}}\left\{ {\begin{array}{*{20}{c}}
a&b&c\\
{d - 1}&e&f\\
g&h&j
\end{array}} \right\} = \left[ {\frac{{{B_d}\left( {ag,fe} \right)}}{{d\left( {d + 1} \right)}} - \frac{{{B_c}\left( {ab,jf} \right)}}{{c\left( {c + 1} \right)}}} \right]\left\{ {\begin{array}{*{20}{c}}
a&b&c\\
d&e&f\\
g&h&j
\end{array}} \right\}\\
{A_q}\left( {pr,st} \right) = {\left[ {\left( { - p + r + q} \right)\left( {p - r + q} \right)\left( {p + r - q + 1} \right)\left( {p + r + q + 1} \right)} \right]^{\frac{1}{2}}}\\
 \times {\left[ {\left( { - s + t + q} \right)\left( {s - t + q} \right)\left( {s + t - q + 1} \right)\left( {s + t + q + 1} \right)} \right]^{\frac{1}{2}}}\\
{B_q}\left( {pr,st} \right) = \left[ {q\left( {q + 1} \right) - p\left( {p + 1} \right) + r\left( {r + 1} \right)} \right]\left[ {q\left( {q + 1} \right) - s\left( {s + 1} \right) + t\left( {t + 1} \right)} \right]
\end{array}
\end{flalign}
Geometrical interpretations of $A$'s as proportional to products of areas of triangular faces and of $B$'s as angular functions of associated structures, will serve for further work on the dynamical description of general spin networks. As noted in \cite{AnAqMa}, Eq. \ref{eq:xyrecur} can be derived by setting $h=0$ in Eq. \ref{eq:recur9j}, and using the property that a $3nj$ symbol downgrades to a $(3n-1)j$ symbol when one of its entries is zero.  In conclusion, expanding the discussion of Eq. 43 in \cite{AnAqMa}, we suggest that the ``screen" for the above $9j$ symbols is three-dimensional, and generalization to higher spin networks should be straight forward.\\ \\
\textbf{Acknowledgement.}  We thank Professor Annalisa Marzuoli for many productive discussions during this research.

\bibliographystyle{splncs}

\bibliography{aquila,sort}

\end{document}